\documentclass[aps,pra,twocolumn,showpacs,superscriptaddress,groupedaddress]{revtex4}

\usepackage[utf8x]{inputenc}
\usepackage{amsmath,amssymb,amsfonts,subfigure,braket,color,dsfont}
\usepackage[english]{babel}
\usepackage{graphicx} 
\usepackage{sistyle} 
\usepackage{listings}
\usepackage{gensymb}

\begin{document} 
 
\title{First-order thermal insensitivity of the frequency of a narrow spectral hole in a crystal}

\author{S. Zhang} 
\affiliation{LNE-SYRTE, Observatoire de Paris, Universit\' e PSL, CNRS, Sorbonne Universit\' e, Paris, France}
\author{S. Seidelin}\email{signe.seidelin@neel.cnrs.fr}
\affiliation{Univ. Grenoble Alpes, CNRS, Grenoble INP and Institut N\' eel, 38000 Grenoble, France}
\author{R. Le Targat} 
\affiliation{LNE-SYRTE, Observatoire de Paris, Universit\' e PSL, CNRS, Sorbonne Universit\' e, Paris, France}
\author{P. Goldner} 
\affiliation{Chimie ParisTech, Universit\' e PSL, CNRS, Institut de Recherche de Chimie Paris, 75005 Paris, France} 
\author{B. Fang} 
\affiliation{LNE-SYRTE, Observatoire de Paris, Universit\' e PSL, CNRS, Sorbonne Universit\' e, Paris, France}
\author{Y. Le Coq}
\affiliation{LNE-SYRTE, Observatoire de Paris, Universit\' e PSL, CNRS, Sorbonne Universit\' e, Paris, France}

\date{\today}

\begin{abstract}
The possibility of generating an narrow spectral hole in a rare-earth doped crystal opens the gateway to a variety of applications, one of which is the realization of an ultrastable laser. As this is achieved by locking in a pre-stabilized laser to the narrow hole, a prerequisite is the elimination of frequency fluctuations of the spectral hole. One potential source of such fluctuations can arise from temperature instabilities. However, when the crystal is surrounded by a buffer gas subject to the same temperature as the crystal, the effect of temperature-induced pressure changes may be used to counterbalance the direct effect of temperature fluctuations. For a particular pressure, it is indeed possible to identify a temperature for which the spectral hole resonant frequency is independent of the first-order thermal fluctuations. Here, we measure frequency shifts as a function of temperature for different values of the pressure of the surrounding buffer gas, and identify the ``magic'' environment within which the spectral hole is largely insensitive to temperature.

\end{abstract}


\maketitle

\section{Introduction}

Rare-earth ions doped inside a crystalline matrix, when utilized at cryogenic temperatures, can exhibit optical features with extremely narrow spectral width, in fact record-narrow, for certain combination of materials, when compared to other solid state materials. Classical~\cite{Berger2016} and quantum~\cite{Nilson2005,Bussieres2014,Walther2015,Maring2017} information processing schemes, quantum memories~\cite{Zhong2017,Laplane2017}, and more recently, quantum opto-mechanics~\cite{Molmer2016,seidelin2019,ohta2021} are domains that take advantage of such properties. Furthermore, high precision laser frequency stabilisation~\cite{Julsgaard2007,Thorpe2011,Gobron2017,galland2020_OL} has also been shown in the last decade, to potentially benefit from the use of such materials. In this latter application, even more than others, the perturbations potentially induced by the cryogenic environment must be sufficiently controlled in order to minimize frequency fluctuations of the narrow optical structures. Such perturbations include vibrations, which influence the spectral hole through variations in mechanical strain~\cite{galland2020}, as well as temperature fluctuations, which typically will give rise to frequency fluctuations of the spectral structure. In general, the frequency of the spectral hole has been observed to be highly dependent on temperature (proportional to $T^4$~\cite{Konz2003}, with $T$ in Kelvin). For instance, europium ions doped into an yttrium ortho-silicate matrix is the material that has been explored so far for optical frequency metrology applications due to the relative ease of obtaining kilo-Hertz narrow spectral features by the technique known as spectral hole burning. In this case, the relative frequency instability of a given spectral hole of the order of $10^{-15}$ requires a temperature instability below $10^{-4}$ K, which is challenging to maintain durably for standard 4~K cryogenic system. It is however possible, when surrounding the crystal in a cryogenic temperature helium buffer-gas which is subject to the same temperature fluctuations as the crystal, to identify a particular pressure of the helium, for which the resulting frequency fluctuations due to the temperature variations are minimized. More precisely, a shift in the temperature gives rise both to a direct shift in frequency, but also to a change in pressure, which in turn also shifts the frequency (through a modification of the mechanical strain of the crystal). For a given pressure and temperature set-point, it is possible to have these two effects compensate each other to the first order, such that only a quadratic temperature dependence remains~\cite{Thorpe2013}, realizing a ``magic'' environment  for minimizing frequency fluctuations around the minimum of the temperature-frequency curve. We will in the following study these temperature effects, and we determine a suitable pair of temperature-pressure values for which such environment is realized.

\section{The experimental system}

The rare-earth ion used in our study is the europium ion ($\rm Eu^{3+}$) doped into a $\rm Y_2SiO_5$ crystal (abbreviated to Eu:YSO in the following) mentioned above. This system exhibits one of the most narrow linewidths observed for an optical transition in a solid state environment. We use a 0.1 at.\% europium doping density. The $^7F_0 \rightarrow$ $^5D_0$ optical transition corresponds to a wavelength of  580\,nm. This particular transition can exhibit coherence times of the order of miliseconds~\cite{Equall1994,Oswald2018}. The europium ion can be substituted in two different positions in the YSO matrix, commonly referred to as site 1 and site 2 (with vacuum wavelengths of 580.04\,nm and 580.21\,nm, respectively). In the following, we will only consider site 1, but we expect that similar results can be obtained for site 2.  Due to the size disparity between $\rm Eu^{3+}$ and $\rm Y^{3+}$ ions, the doping causes deformations in the crystalline matrix, that in turn gives rise to an inhomogeneous broadening of the absorption profile of $\rm Eu^{3+}$ ions with a linewidth typically of the order of 2~GHz in full width half maximum (FWHM) for 0.1\% doping level. To take advantage of the narrow single ion linewidth, and still exploit a large fraction of the ions, we employ the techniques of spectral hole burning~\cite{Konz2003}, which is essentially optically pumping a subclass of ions, identified by their common resonance frequency, away to a metastable state, leaving a `hole' in the absorption spectrum. The spectral holes reproduce the narrow features of the single ions, but posses a much larger signal-to-noise ratio. We can achieve experimentally spectral holes of the order of 3-4 kHz FWHM. The experimental setup is similar to the one used in our previous studies (see for instance~\cite{galland2020,zhang_2020a,zhang_2020b}) and the reader is referred to these papers for experimental details on the laser and photo-detection system. In particular, the measurements require an ultra-stable laser, which is realized by frequency-locking a laser to a reference cavity using the Pound-Drever-Hall locking scheme. We obtain a laser frequency instability close to $10^{-14}$ for time constants in the range of 1-100\,s, and a corresponding linewidth of a few Hertz. The absorption profiles, obtained by first burning a spectral hole with the laser at comparatively large optical power ($ \approx 10 ~\mu$W), and later scan the frequency of the same laser but operated at much lower power ($ \approx 100$~nW), are recorded using avalanche photodiodes and the heterodyne method presented in~\cite{galland2020}.

\begin{figure}[t]
\centering
\includegraphics[width=80mm]{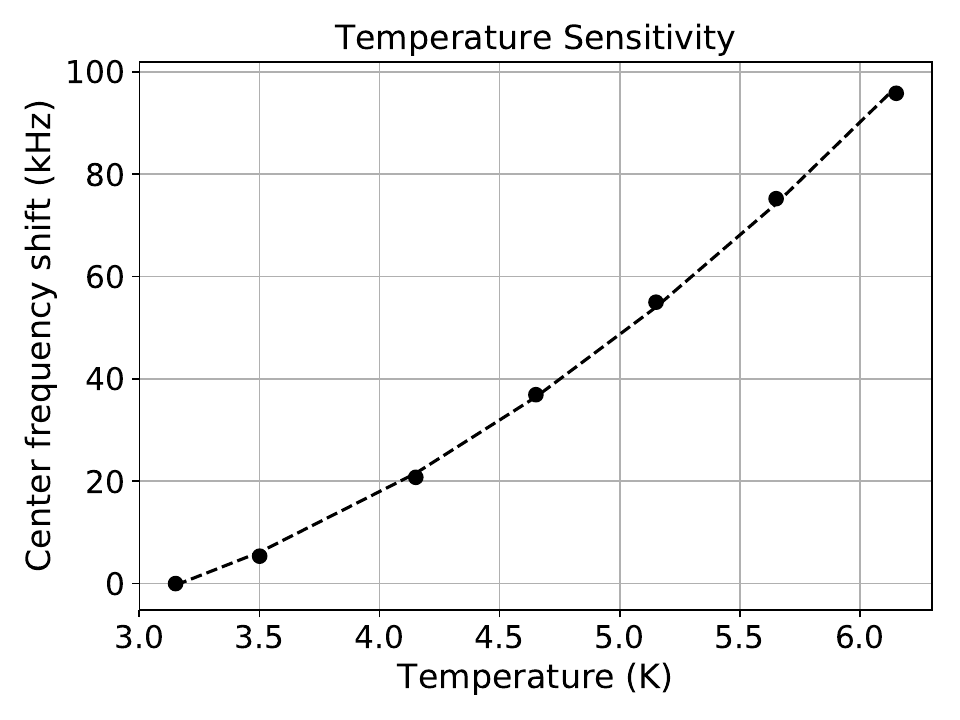}
\caption{\label{TP} The thermal sensitivity of the spectral hole (crystal site 1) in a vacuum chamber. The crystal is attached to a cold copper plate using silver lacquer. The data points represents the measured frequency shift of the spectral hole at steady state when the temperature is changed and stabilized, with a starting temperature at 3.15 K. The statistical error-bars (for both temperature and frequency) are of the size of the data markers, and thus appear invisible. The dashed line is the fitting of the data with a quadratic model.}
\end{figure}

We use a commercially available, low-vibrations, closed cooling-cycle cryostat (OptiDry from MyCryoFirm), to maintain the crystal (cuboid $8\times8$ mm transverse dimensions, probed along the 4mm long b crystalline axis with polarization along the D1 dielectric axis) at 3-6\,K. Optical access is ensured by BK7 view ports with more than 1~cm clearance diameter. To confirm the dependence of frequency of the spectral hole on temperature, our first series of measurements are performed in vacuum (pressure $<$ 10$^{-6}$ Pa), and the temperature is controlled by positioning the crystal directly on a copper plate, in thermal contact with the cold plate of the cryostat but equipped with electrical heaters and thermistor controlled by a commercial active servo-loop specialized for cryogenic systems (Lakeshore 350). We note that the temperature is stabilized to better than $1$~mK at each measurement point, see fig.~\ref{temp4K} and discussions below.

Finally, also note that the frequency reference is provided by the reference cavity mentioned above, whose own frequency drift is always lower than 50\,Hz (in absolute value) over the course of the full measurement runs (a few minutes each) and thus deemed negligible. This has been verified by an independent measurement against a metrological composite reference utilizing an active hydrogen maser, an optical frequency comb and an optical reference cavity at 1.5~$\mu$m wavelengths.

%Finally, also note that the frequency reference is provided by the reference cavity mentioned above, whose own frequency drift, which is always lower than 50\,Hz (in absolute value) over the course of full measurement runs (as measured independently against a metrological composite reference utilizing an active hydrogen maser, an optical frequency comb and an optical reference cavity at 1.5~$\mu$m wavelengths) and is thus deemed negligible for the couple of minutes of duration of each measurement run}. 

\section{Measurement results}

The results of these measurements are shown in fig.~\ref{TP}, which represents the measured frequency shift of the spectral hole at different temperatures.

\begin{figure}[t]
\centering
\includegraphics[width=85mm]{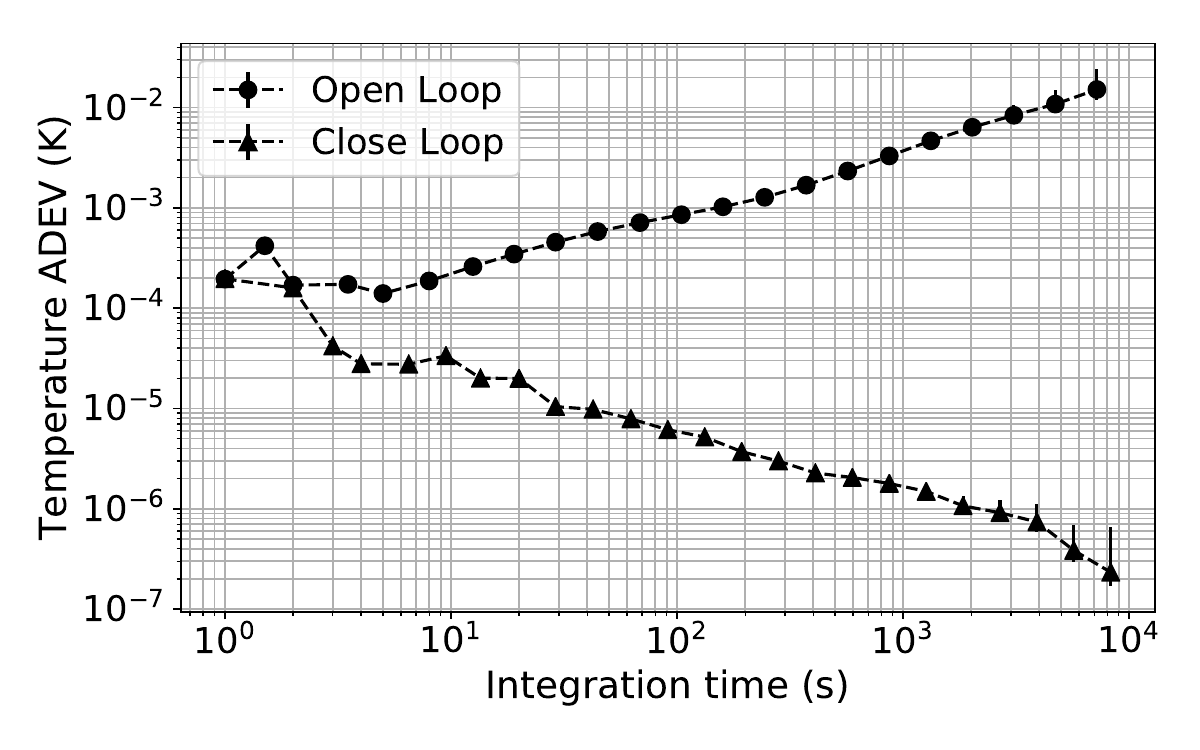}
\caption{\label{temp4K} Typical temperature fluctuations (Allan deviation) as a function of averaging (integration) time. The circles show the free running temperature, and the triangles show the stabilized temperature (as measured by the sensor used in the servo-loop, since we do not currently have the necessary extra sensors for an independent out-of-loop measurement). The temperature fluctuations are on the order of hundreds of $\mu$K or below for an averaging time of $1$~s or longer once the servo-loop is turned on.}
\end{figure}

Figure~\ref{temp4K} shows typical temperature fluctuations of the $4$~K stage of our cryostat as a function of averaging time. The temperature is measured by a Carbon Ceramic Sensor$^{\rm TM}$ (Temati) calibrated by the manufacturer of our cryostat. Both free running and stabilized  temperatures are shown. Without a servo-loop, we observe a slow drift at long time scales. Once the stabilization is active, the temperature fluctuations are on the order of hundreds of $\mu$K or below for an averaging time of $1$~s or longer.

For the crystal in vacuum, for ions in site 1, we measure a thermal sensitivity of 17 $\pm$ 1 kHz/K at 3.15 K, corresponding to a linearization of the fitting curve shown in fig.~\ref{TP}. This result is situated in between measurements performed by two different groups, yielding in one case 21 $\pm$ 2 kHz/K~\cite{Konz2003} and in the other case 10 kHz/K~\cite{Thorpe2013}. These variations might be attributed to differences in the exact mounting of the crystal. In our case, in the vacuum configuration, the crystal has been glued to the support for better thermal contact, which might alter the effective sensitivity. This is contrary to what we do in the following when the crystal is immersed in a helium gas, where it is positioned directly on the support without glue, and thus can freely expand and contract.

%Note however that it is outside the error bar of the value given earlier by a third group~\cite{Konz2003} of 21 $\pm$ 2 kHz/K, although this last value was extrapolated from higher temperature measurements, and potentially less accurate around 3~K.}

\begin{figure}[t]
\centering
\includegraphics[width=80mm]{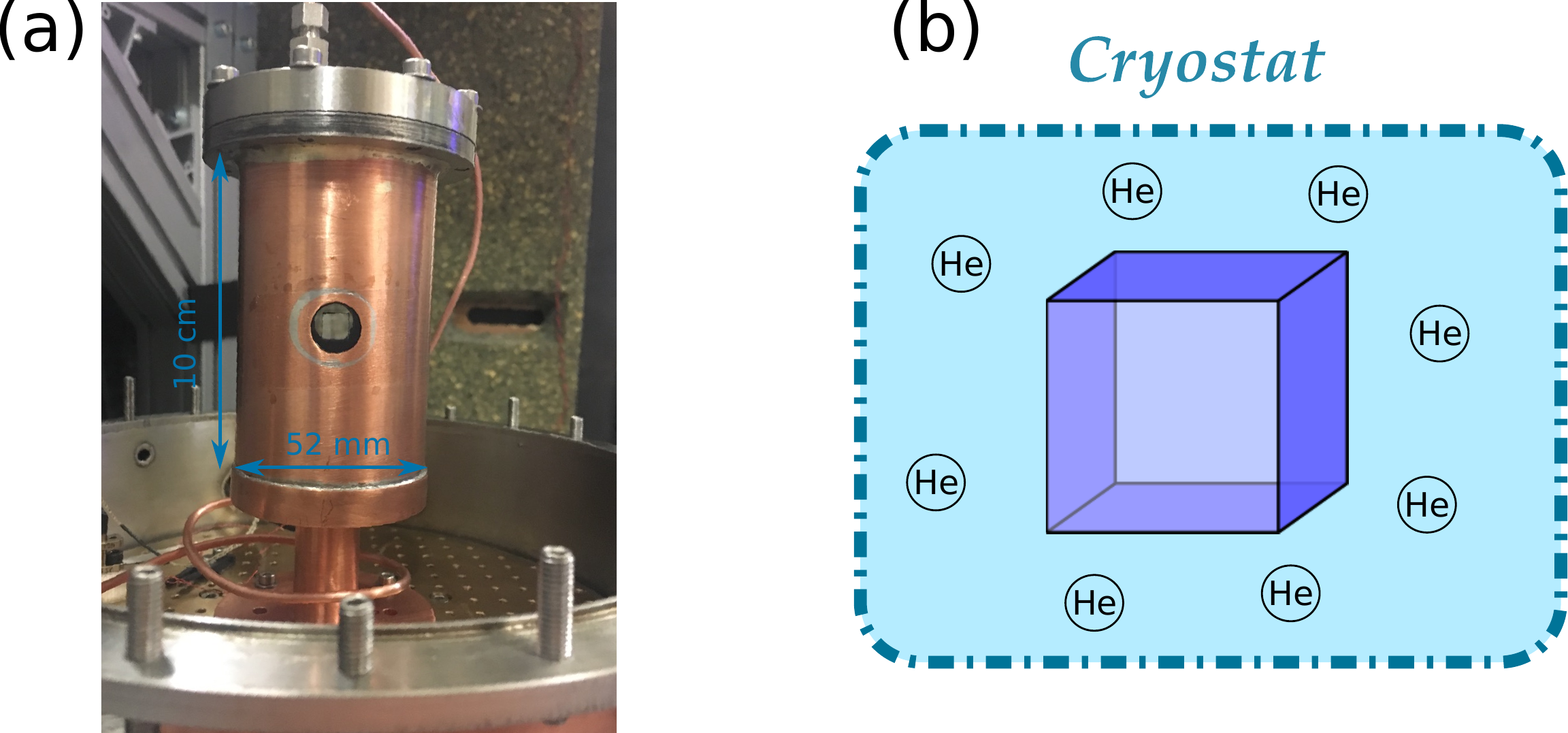}
\caption{\label{IVC} The internal vacuum chamber for achieving the temperature and pressure coupled environment. A photo of the IVC which consists of a cylindrical cell, with a diameter of 52 mm, is shown in (a), in which a tube allows the injection of helium gas, forming an isotropic pressure on the crystal as illustrated in (b).}
\end{figure}  
  
In order to identify and realize a ``magic'' environment, we need to be able to create a controlled, isotropic pressure around the crystal, with a temperature dependence that follows that of the crystal and which induces pressure changes similar to that of an ideal gas at (mostly) fixed volume and atom number. Therefore, we need a sealed environment, which conserves the initial amount of gas (maintaining a constant number of gas particles) such that a change in temperature gives rise to a change in pressure. To do so, we add a smaller, internal vacuum chamber (IVC) of approximately $2\times 10^{-4}$ m$^3$, in which we can fill a variable amount of helium gas, see fig.~\ref{IVC}. On the top of the IVC, a pipe connecting the inside of the cell to the exterior of the cryostat allows for either gas injection or for pumping of the cell through two manual valves situated at room temperature. During our measurements, both valves are closed such that the total amount of injected helium gas is constant. The constant volume is also achieved due to rigid tubings in addition to the IVC itself. A pressure gauge (CERAVAC CTR100, 0.2 \% accuracy) is positioned close to the injection point at room temperature. Here, instead of a copper plate used for the measurements in vacuum, the crystal is now put on a plastic post (which is thermally isolating) inside the IVC, such that the temperature of the crystal is determined by the temperature of the surrounding gas. The temperature stability of the IVC is similar to the one obtained without the IVC shown in fig.~\ref{temp4K}. Note that we do not have direct information about the temperature of the crystal itself as the IVC is not equipped with leak-proof electrical connections.

\begin{figure}[t] 
\includegraphics[width=80mm]{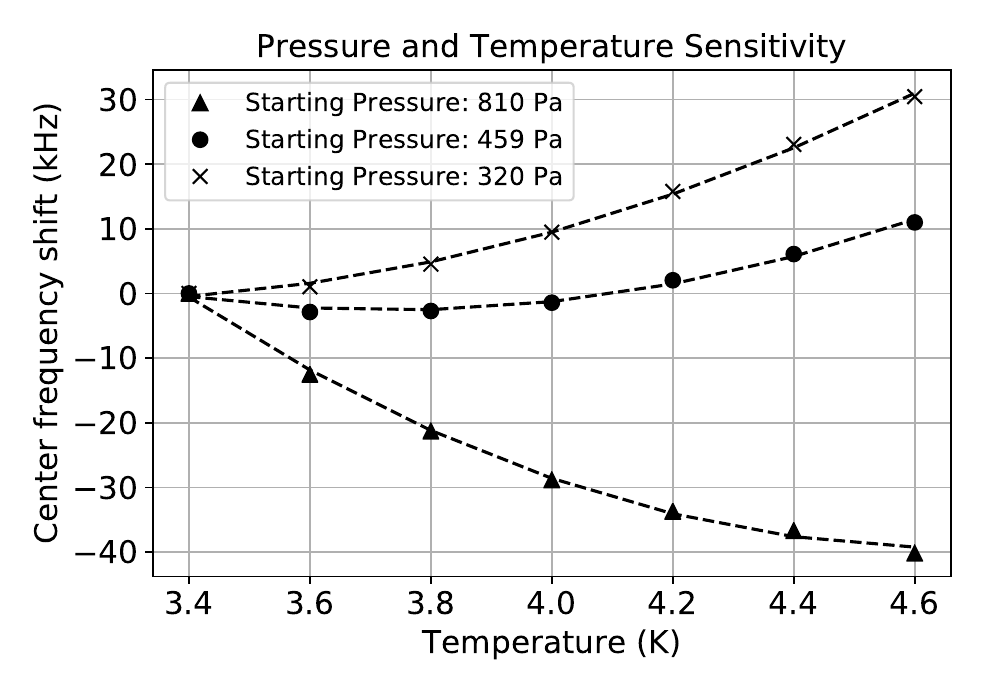}
\caption{\label{freqvstemp}  When temperature is varied, the spectral holes shift in frequency. Data are shown for three different initial pressures corresponding to a starting temperature of 3.4 K for crystal site 1. When the temperature is increased, the pressure (not shown) is also led to evolve (increase in fact). Each data set is fitted using a quadratic model (dashed lines). In the middle curve, indicated by circle shaped markers, a magic point at 3.7 K is identified, for which a first-order cancellation of temperature and pressure effects is observed. The statistical error-bars (for temperature and frequency) are also included here. Since they are of the size of the data markers, they appear to be invisible.}
\end{figure}   

The measurement sequence is given as follows: we first pump the IVC down to vacuum, then add in helium gas to reach the desired isotropic pressure. We subsequently measure the frequency of the spectral hole at different temperature set points, starting at 3.4~K. Before each measurement, we allow the crystal to thermalize with the surrounding gas (approximately 2 minutes, depending on the pressure). Thermalization is confirmed by the stability in time of the frequency of the spectral hole. The measured frequency shift of the spectral holes as a function of temperature are shown in figure~\ref{freqvstemp}. In particular, one magic environment can be observed for the middle curve, with an initial pressure of 459 Pa (circle markers). This curve presents a minimum at 3.7~K, corresponding to a pressure of 526 Pa. This pressure value corresponds to a direct measurement, and deviates slightly from the value obtained by assuming that it follows the ideal gas law, in particular due to a temperature gradient along the tubings. The data can be fitted with a quadratic model (residual temperature dependence at the magic environment), arising from second order development of the theoretical expression $\delta \nu= \alpha(T_0+\delta T)^4 +\beta (T_0+\delta T)-\nu_0$, where $T_0$ is set to the initial temperature of 3.4\,K for which spectral hole imprinting occurs, $\delta \nu$ is the frequency offset of the hole with respect to its value at $T_0$ ($\nu_0$ therefore corresponds to the hypothetical frequency of the hole which would be obtained at 0\,K in this model), $\delta T$ is the offset temperature from the $T_0$ value, $\alpha$ arises from the direct temperature dependence of the crystal and $\beta$ from the pressure-induced shift (with the pressure itself assumed to be varying linearly with temperature \cite{Thorpe2013} over the explored temperature range). The obtained second order sensitivity is of the order of 20 kHz/K$^{-2}$. In the perspective of realizing an ultra-stable laser by locking to a spectral hole, at this magic point, typical temperature fluctuations of the order of $10^{-4}$~K (as discussed above) imply a fractional frequency instability of the laser of $4\times10^{-19}$, which is well below the instability resulting from other sources of noise currently identified on our experiment, and thus no longer a limiting factor. We emphasize however that this cancellation of temperature sensitivity is valid only provided that the state of thermal equilibrium between the crystal, helium gas and mechanical enclosure is reached, which takes a finite time and may not allow perfect cancellation at short timescales. Detailed time response analysis to various sources of temperature perturbation, beyond the scope of this work, may be necessary to assess the efficiency of this method at short timescales. 

We point out that the parameters defining the observed magic environments (starting pressure, and temperature at which the minimum is achieved) should be the same (for the same transitions) for different samples of europium ions doped into high-purity YSO single crystals to within the measurement uncertainty of these values (an uncertainty in pressure of 0.2 \% corresponding to 1 Pa for the 459 Pa curve in fig.~\ref{freqvstemp}, and below 1 millikelvin for all temperature measurements). This is due to the fact that the europium ions have the same valence and approximately the same ionic radius as the yttrium atoms they substitute, thus the observed pressure and temperature sensitivities (and their combination) vary minimally from one sample to another.

\begin{figure}[t]
\centering
\includegraphics[width=80mm]{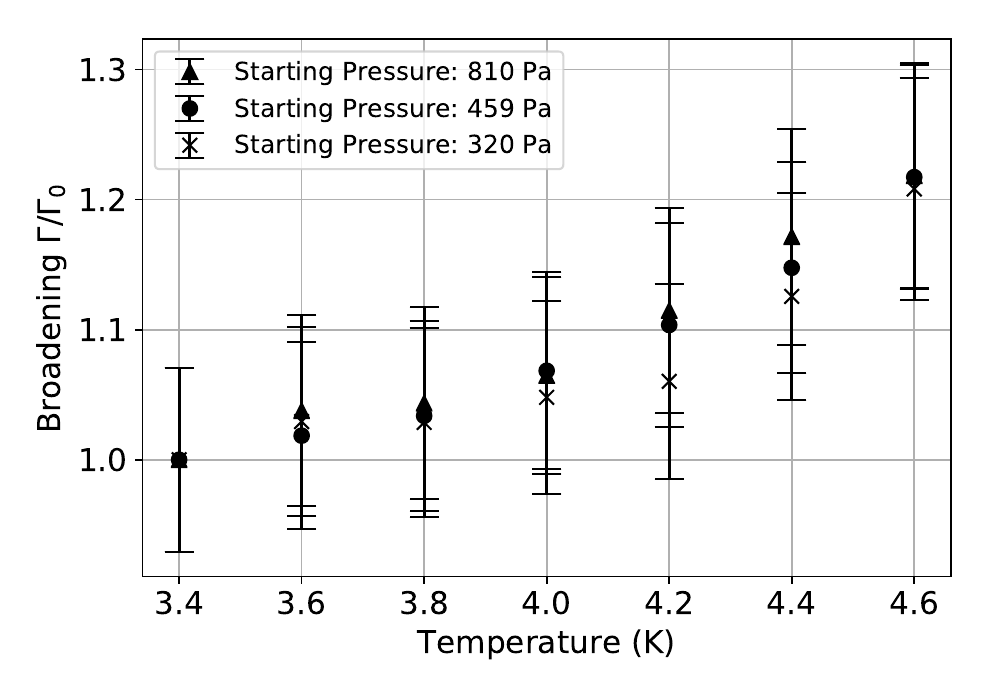}
\caption{\label{width} The normalized FWHM of the spectral holes (with respect to the initial spectral hole) as a function of temperature. The values correspond to the measurements presented in fig.~\ref{freqvstemp}.}
\end{figure}

In addition to a shift in the center frequency of the spectral holes, we also plot the effect of the temperature on the widths of the spectral holes, normalized to the initial width, see fig.~\ref{width}.  In the context of achieving an ultrastable laser by locking its frequency to a narrow spectral hole, the increase in linewidth does not in principle affect the central frequency, but can change the response of the system (and therefore the effective gain of the servo-loop) and potentially diminish the advantages of the narrowness of the structure, depending on the extent to which the hole is broadened. Such effects therefore deserve further investigation. Using a Lorentzian model, we extract the FWHM of the spectral holes from the measurement presented in fig.~\ref{freqvstemp}. The quadratic sum of the fitting error and 5\% of the FWHM (a conservative estimate of the effect of a slight asymmetry of the spectral hole profiles) is used to evaluate the uncertainty for each FWHM. We verified independently that the broadening related to repeated scans of the spectral hole is negligible. Since the initial widths are slightly different for each different starting pressure (5.80 kHz for 810 Pa, 6.12 kHz for 459 Pa and 5.82 kHz for 320 Pa) due to variations in the experimental conditions during the initial hole burning process, a simple normalization with respect to the initial width of the spectral hole is carried out. The errorbar on the ratio $\Gamma/\Gamma_0$ plotted in fig.~\ref{width} then takes into account that of the initial width and that of the width used for normalization. For each data set, when the temperature is increased, a broadening of up to 20\% can be observed. In particular, in the vicinity of the magic point (3.6 to 3.8 K), a linearization gives a broadening of approximately 0.46 kHz/K, corresponding to 8\% broadening per K. In addition, the three data sets coincide within the errorbars of the measurements, possibly indicating that there could be a common underlying physical mechanism that governs the broadening phenomena, which is not yet fully understood, and may require further investigation and modeling. Nevertheless, the impact of the observed increase of the width in the vicinity of the magic point is negligible in the context of the realization of an ultrastable laser. Since the servo-loop provides a temperature stabilization better than 1 mK, corresponding to $10^{-5}$ relative change in the spectral hole linewidth, we expect a relative change in the effective gain in the laser servo-lock to be of the same order of magnitude, which is negligible.

\section{Discussion and outlook}

In this article, we have here investigated the temperature dependence of a spectral hole, which is of importance in the context of frequency stabilization by locking to a narrow spectral hole. By identifying a ``magic point'' for which the frequency shift of a spectral hole due to direct temperature fluctuations and pressure fluctuations (arising from temperature fluctuations inside a sealed environment) to first order compensate each other, the laser stabilization will not be limited by such fluctuations. We have also provided evidence that the broadening of the spectral hole due to temperature fluctuations around this magic point are sufficiently small as not to influence the realization of an ultrastable laser using this technique. As mentioned above, this statement will hold true as long as other sources of error maintain the instability of the achieved ultrastable laser above $4\times10^{-19}$. This limit is well below what can currently be achieved with rare-earth doped systems, due to the presence of other noise sources such as mechanical vibrations and fluctuating surrounding fields, as well as detection noise~\cite{galland2020_OL}. It is also well below what is currently achieved with cryogenic Fabry-Perot cavity based laser stabilization systems~\cite{Matei_2017,Robinson_2019}. More importantly, the need for ultrastable lasers with instabilities below this value is fairly limited, as their main utility is to probe atomic clocks which are currently at best at the $10^{-17}$ near 1\,s timescale~\cite{Nicholson_2012}. Furthermore, even the use of an optical frequency comb that is required to transfer the spectral purity from the rare-earth-doped stabilized laser to a wavelength of interest is currently limited to a transfer at a stability level of a few 10$^{-18}$ at 1\,s timescale~\cite{Nicolodi_2014}.

The demonstrated technique of temperature stabilization should in principle not be limited to Eu:YSO, but could be applicable in other rare-earth ion/crystal matrix combination allowing for persistent spectral holes, and which are sufficiently narrow in order to being limited in frequency stability by temperature fluctuations. However, as the frequency shifts due to temperature \cite{Kushida_1969} and pressure~\cite{Louchet-Chauvet_2019} exhibit distinct behaviours in different systems, the viability of the scheme must be assessed in each individual case.

%This is due to the fact that the pressure- and temperature sensitivities are comparable in different rare-earth systems, and thus allow for similar cancellation effects. In particular, Erbium ions at telecom wavelengths (for instance $\rm Er^{3+}$:$\rm D^{-}$:$\rm CaF_2)$ which have also been used for frequency stabilization techniques~\cite{Bottger2003}, could possibly benefit from a similar scheme}.

An alternative approach to the demonstrated technique would be to operate laser stabilization at much lower temperature (below 100 mK). When approaching a zero temperature, the slope and curvature of the $T^4$ dependence (assuming no other frequency shifting mechanism occurs at such low temperature) correspondingly decreases, making the temperature fluctuations progressively less influential. For instance, at 100 mK, the linearized coefficient would be of the order of 0.3 Hz/K. For temperature fluctuations of the order 30 $\micro$K at 1\,s, readily achievable at this temperature in a standard dilution refrigerator (in contrast to a standard 4 K cryogenic pulsetube system in which even a millikelvin stability is challenging to reach), this would result in a relative laser instability of $10^{-20}$, significantly below any current limitations. This approach requires however a dilution refrigerator system, adding complexity to the experimental setup, compared to a standard 4~K cryostat used in this work.

\section{Acknowledgements}

We thank Anne Louchet-Chauvet and Thierry Chaneli\`{e}re for useful discussions. The project has received financial support from Ville de Paris Emergence Program, the R\'{e}gion Ile de France DIM C'nano and SIRTEQ, the LABEX Cluster of Excellence FIRST-TF (ANR-10 LABX-48-01) within the Program ``Investissements d'Avenir'' operated by the French National Research Agency (ANR), the CNRS through the MITI interdisciplinary programs, and the 15SIB03 OC18 and 20FUN08 NEXTLASERS projects from the EMPIR program co-financed by the Participating States and from the European Union's Horizon 2020 research and innovation program.

\end{document}